\documentclass[10pt,
aps,pra,
reprint,
notitlepage,
superscriptaddress,
frontmatterverbose, 
showpacs,
preprintnumbers,nofootinbib,
amsmath,amssymb,
]{revtex4-1}
\usepackage[version=3]{mhchem} 
\usepackage[utf8x]{inputenc}
\usepackage{graphicx}
\usepackage[english,british]{babel}
\usepackage{ifpdf}
\usepackage{natbib}
\usepackage{multirow}
\usepackage{appendix}
\newcommand{\centric}{$\text{Im}\bar{3}\text{m}$ }
\newcommand{\acentric}{$\text{I}\bar{4}3\text{m}$ }

\newcommand{\AAA}{$\text{\AA~}$}
\newcommand\vs{\textit{vs.~}}
\newcommand{\p}{$^{+}$}
\newcommand{\pp}{$^{2+}$}
\newcommand{\na}{Na\p~}
\newcommand{\ca}{Ca\pp~} 
\newcommand{\po}{K\p~}
\newcommand{\sr}{Sr\pp~}
\newcommand{\li}{Li\p~}
\begin{document}
\bibliographystyle{unsrtnat}
\preprint{}
\title{Understanding nanopore window distortions \\in the reversible molecular valve zeolite RHO}
\author{Salvador \surname{R. G. Balestra}}
\affiliation{Área de Química--Física, Universidad Pablo de Olavide, Ctra. Utrera km 1, 41013 Seville, Spain}
\author{Said \surname{Hamad}}
\email{(SH) \url{said@upo.es} and (SC) \url{scalero@upo.es}}
\affiliation{Área de Química--Física, Universidad Pablo de Olavide, Ctra. Utrera km 1, 41013 Seville, Spain}
\author{A. Rabdel \surname{Ruiz-Salvador}}
\affiliation{Área de Química--Física, Universidad Pablo de Olavide, Ctra. Utrera km 1, 41013 Seville, Spain}
\author{Virginia \surname{Dom\'{i}nguez--Garc\'{i}a}}
\affiliation{Departamento de Electromagnetismo y Física de la Materia, and Instituto Carlos I de Física Teórica y Computacional Universidad de Granada, 18071 Granada, Spain}
\author{Patrick J. Merkling}
\affiliation{Área de Química--Física, Universidad Pablo de Olavide, Ctra. Utrera km 1, 41013 Seville, Spain}
\author{David \surname{Dubbeldam}}
\affiliation{Van't Hoff Institute for Molecular Sciences, University of Amsterdam, Science Park 904, 1098 XH Amsterdam, }
\author{Sof\'ia \surname{Calero}$^{~1,*}$}
\homepage[Visit: ]{http://www.upo.es/raspa/}
\date{\today}
\begin{abstract}
Molecular valves are nanostructured materials that are becoming popular, due to their potential use in bio-medical applications. However, little is known concerning their performance when dealing with small molecules, which are of interest in energy and environmental areas. It has been observed experimentally that zeolite RHO shows unique pore deformations upon changes in hydration, cation siting, cation type, and/or temperature-pressure conditions. By varying the level of distortion of double 8-rings it is possible to control the adsorption properties, which confers a molecular valve behavior to this material. We have employed interatomic potentials-based simulations to obtain a detailed atomistic view of the structural distortion mechanisms of zeolite RHO, in contrast with the averaged and space group restricted information that can be retrieved from diffraction studies. We have modeled the pure silica zeolite RHO as well as four aluminosilicate structures, containing \li, \na, \po, \ca and \sr cations. It has been found that the distortions of the three zeolite rings are coupled, although the four-membered rings are rather rigid and both six- and eight-membered rings are largely flexible. A large dependence on the polarizing power of the extra-framework cations and with the loading of water has been found for the minimum aperture of the eight-membered rings that control the nanovalve effect. The energy barriers needed to move the cations across the eight-membered rings are calculated to be very high, which explains the origin of the experimentally observed slow kinetics of the phase transition, as well as the appearance of metastable phases.
\end{abstract}
\maketitle
\section{Introduction}
Molecular valves are a class of molecular devices that allow molecular transport in a controlled way through gate opening or trapdoor mechanisms. Valve action is typically performed by a molecule that is attached to the material, either by covalent bonds, hydrogen bonds or supramolecular interaction. In presence of external stimuli, such as temperature, pressure, pH, molecular or ion chemical potential, this molecule is able to change its configuration to allow the molecular flow. This ability has attracted huge attention during the last years, due to its impact in delivering medium and large size active molecules for medical applications \cite{mengvalve2010,C1MD00158B,doi:10.1021/nn3018365,doi:10.1021/nn101499d}.
\begin{figure}[h!]
	\begin{center}
	   \centering
	   \includegraphics[width=0.48\textwidth]{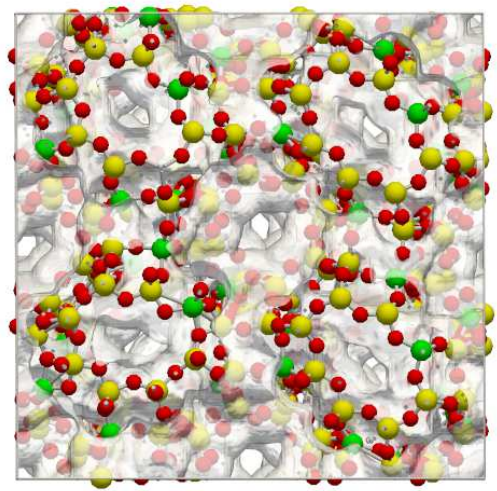}
	   \caption{\label{fig:i43m}
	   Snapshot of a distorted form of RHO-type zeolite obtained by Molecular Dynamics. An isoenergy surface is shown in translucent gray. Extra-framework cations are omitted for clarity.
	   }
	\end{center}
\end{figure}
However, for small molecules, such as carbon dioxide or small hydrocarbons, operating the molecular valve through an attached molecule is less suitable, as a higher structural control is required for this kind of molecules. In this context, zeolites and metal-organic frameworks (MOFs) appear as candidate materials, due to their crystalline nature and their smaller nanopore windows. As zeolites and MOFs exhibit molecular sieving properties, the conjunction of these properties with particular structural flexibility can give rise to molecular valve behavior.

\citeauthor{titanosilicate_adsorption_nature_2001}\cite{titanosilicate_adsorption_nature_2001} were pioneering in showing that efficient separation of small molecules can be achieved by exploiting framework flexibility. In these selected frameworks, pore window diameters can be tuned by means of temperature to separate \ce{O2}:\ce{N2}, \ce{N2}:\ce{CH4}, and \ce{CH4}:\ce{C2H6} mixtures. Zeolite RHO has a reversible operating mechanism, which can be controlled to smoothly switch from one stable form to another, by exposure to vacuum, dehydration, and changes in temperature \cite{RHO_divalent_cation_jacs,RHO_cation_siting,RHO_relocations_cations_reisner,RHO_dehydration_jpc,RHO_pressure_yongjaeLee,RHO_temperature_corbin84_jpc,RHO_Co2_palomino}.   
 Its potential use in applications that make use of the molecular sieving of small molecules is huge, making this zeolite an excellent case for the study of thermally resistant and highly flexible materials (see Figure \ref{fig:i43m}). Indeed, \citeauthor{lozinska_RHO_co2_jacs_2012} studied recently \ce{CO2} adsorption and separation in several forms of univalent metal-exchanged zeolite RHO \cite{lozinska_RHO_co2_jacs_2012}, and found that the Na-form is the best candidate for applications in selective kinetic gas separation of \ce{CO2}. They performed adsorption experiments, combined with crystallographic analysis, which showed that the observed molecular sieving behavior is due to a molecular valve effect associated to the extra-framework cations that control the molecular passage through windows between cages. This is a key aspect, involved in the high selectivity of this material for \ce{CO2}:\ce{CH4} separation. In another recent paper, \citeauthor{RHO_Co2_palomino}\cite{RHO_Co2_palomino} showed that in a mixed Na-Cs zeolite RHO, a reversible phase transition from \acentric to \centric, and vice versa, can be driven by the presence of adsorbed \ce{CO2} molecules. They found that this zeolite exhibits the highest selectivity towards \ce{CO2}:\ce{CH4} separation, which was related to the polarity of the framework (ratio \ce{Si}/\ce{Al} = 4.5) and the pore opening due to the phase transition triggered by \ce{CO2} adsorption.
\begin{figure}[h]
	\begin{center}
	   \centering
	   \includegraphics[width=0.5\textwidth]{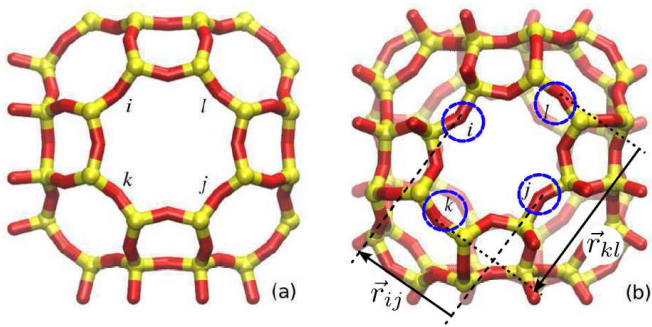}
	   \caption{\label{fig:distortion_parameter}
	   Zeolite RHO in the \emph{a}) \centric and \emph{b}) \acentric space groups. $r_{ij}$, and $r_{kl}$ are used for measuring the distortion parameter $\Delta$ as defined in equation \ref{Deltadef}.
	   }
	\end{center}
\end{figure}

The aluminosilicate zeolite RHO, which adopts an \centric centric structure for temperatures above 300~K, undergoes a significant pore distortion at lower temperatures, resulting in the stabilization of the \acentric acentric form (Figure \ref{fig:distortion_parameter}b). In order to quantify the degree of distortion, \citeauthor{RHO_delta_Parise_1983} \cite{RHO_delta_Parise_1983} introduced a measure that relates the average value of the degree of ellipticity of the double 8-rings (D8R), describing the pore opening windows, as shown in equation \ref{Deltadef}.
\begin{equation}
   	\label{Deltadef}
    \Delta = \frac{1}{2}\langle\lvert r_{ij}-r_{kl}\rvert\rangle
\end{equation}
where $r_{ij}=\lVert\vec{r_i}-\vec{r_j}\rVert $ and $i$, $j$, $k$, $l$ are defined in Figure \ref{fig:distortion_parameter}.

It has been found that zeolite RHO, upon insertion of divalent cations, undergoes an abrupt structural change towards the acentric form \cite{RHO_divalent_cation_jacs}, with 2-6\% changes in the cell parameters and $\Delta$ values $\sim 2$ \AAA in in the acentric form. This is in contrast with the smoother behavior found in the monovalent exchanged zeolite RHO, which exhibits changes in the cell parameters $\sim 3$ \% lower and has lower values of the $\Delta$ parameter (circa 1.5 \AA) \cite{RHO_temperature_corbin84_jpc}. It is worth noting that the appearance of framework distortions in deuterated zeolite RHO \cite{RHO_dehydration_jpc} suggests that this behavior is inherent to the framework topology and that the extra-framework cations act as modulators. This observation is in agreement with our previous paper \cite{Balestra2013}, in which we showed, by means of interatomic potential-based energy minimizations, that pure silica zeolite RHO has two stable forms (with $\Delta=0$ and $\Delta\neq  0$) and it could experience a phase transition between them (from \acentric to \centric space group) at high temperatures. This result was supported by the calculations of X-ray Diffraction patterns.

The structural picture discussed above, which is also used to explain the molecular sieving properties of the mentioned works, rests on the averaged and space-group restricted information retrieved by diffraction techniques. Computer modeling offers the opportunity to overcome these limitations and hence to provide a deeper insight into the distortion mechanisms. In the present context this is of particular relevance, since the mechanism involved in the molecular valve behavior in zeolites and related materials is still poorly understood.

In this work we carry out a detailed atomistic simulation study in order to gain a better understanding of the relationship between the nature of the extra-framework cations and the flexibility of the material, taking into account the concomitant impact on molecular sieving properties. Existing approaches for accurately modeling cation location and/or large variations of cell volumes in zeolites consider either poorly deformable frameworks \cite{n202mellot,vitale1997,maurin2005,plant2006}, or those having low ratios between the cation number and their potentially accessible sites \cite{ruiz1998aluminum,ruiz1999silicon,B102231H}. For all these reasons, in the early stage of this work we were unsuccessful in reproducing the crystal structure of several cation-forms of zeolite RHO by applying these methods. For the present study we have then developed a methodology for accurately modeling the structural behavior of crystalline nanoporous solids undergoing large structural variations, which combines cycles of Monte Carlo (MC) calculations, energy minimization (EM) and Molecular Dynamics (MD) simulations. In order to study the effect of the ionic polarizing power \textit{i.e.} the ionic charge/size ratio, five metal-forms (Na$^+$, Li$^+$, K$^+$, Ca$^{2+}$ and Sr$^{2+}$) of zeolite RHO were considered. The behavior of the pore windows deserves particular attention, as it is expected to be strongly associated with the nanovalve behavior linked to three main phenomena, namely adsorption, separation and transport properties.

\section{Computational Section}
The methodology used here for the realistic structural modeling of zeolite RHO is inspired by the experience in modeling massively defective ionic solids \cite{A908622F,PSSC200461708}, in which large structural deformations occur. Since one of the results of this research has been the development of the new methodology, it is presented in the next section, along with the rest of the results. As anticipated above, several modeling cycles will be carried out, and therefore the use of first principles calculations is hindered by its CPU cost. Thus, interatomic potentials-based simulations are the method of choice. For the case of zeolite RHO, in particular to model the acentric phase, we noted that the shell model is required, in order to take into account the polarizability of the framework oxygen atoms. To check the importance of polarizability on the structure, several trial configurations of metal-forms of zeolite RHO having acentric \acentric structures were modeled. While the acentric structure was kept during energy minimization, it changed to centric \centric when the shell constant is gradually increased to reach very high values. These high values of the shell constant would make the atoms to behave as non-polarizable atoms. In addition, we used the rigid-ion models of \citeauthor{van1990force} \cite{van1990force,kramer1991interatomic} and \citeauthor{Naseem_potential} \cite{Naseem_potential}, for which the acentric structure is not stable either. Then, the well-known shell-model potentials of \citeauthor{C39840001271} \cite{C39840001271} was used, as they provide accurate structures of complex zeolites. In connection with it, Li$^{+}$, Na$^+$, K$^+$, Ca$^{2+}$, and Sr$^{2+}$ cations were selected considering the existence of good quality interatomic potentials \cite{catlow_potential_1987,catlow_potential_1988,kuvcera2003coordination} that are compatible with the \citeauthor{C39840001271} potentials.

At the experimental synthesis temperature RHO zeolites adopt the centric cubic \centric structure, with only one distinct tetrahedral site in the asymmetric unit cell \cite{syntesis_RHO_1973,mcCusker_1984}, which strongly suggests that a particular ordering of the Al atoms is not likely to occur. In addition, RHO zeolites are synthesized with monovalent cations, usually mixtures of Na$^+$ and Cs$^+$. There is no need to introduce the close pairs Al-Al that would be required if divalent cations were used \cite{Gomez-Hortiguela2010}. Therefore, like in the case of FAU zeolites \cite{dempsey_1969}, Al atoms in RHO zeolites are expected to stay as far as possible from each other, following Dempsey’s rule \cite{dempsey_1969}, as was recently shown to be expected to hold in zeolites, unless framework anisotropy perturbs this behavior \cite{RuizSalvador2013330}. In order to analyze the influence of Al location in the unit cell, we constructed two different configurations, distributing the Al atoms as follows: a) as far as possible in a $2\times 2\times 2$ supercell, and \emph{b}) as far as possible in a single unit cell, which was subsequently expanded to a $2\times 2\times 2$ supercell.

Zeolites are known to contract upon dehydration, which in part is due to the loss of space filling molecules, but more importantly to the increased electrostatic attraction of the extra-framework cations (EFCs) \cite{B607030B}. In zeolite RHO experimental results have shown that the cell contraction caused by dehydration is accompanied by a phase transition from centric to acentric form \cite{RHO_delta_Parise_1983,lozinska_RHO_co2_ChemMat_2014,RHO_relocations_cations_reisner}. Then, with the aim of determining whether the cell contraction \textit{per se} causes the phase transition of the RHO zeolite framework, we modeled the effect of the externally applied pressure on the pure silica structure. Previous work has shown that the chosen potentials perform very well on aluminosilicate zeolites subject to high pressures \cite{white2004pressure}. This is also a test for the applicability of the potentials to model the lower acentric symmetry of zeolite RHO.

\citeauthor{lozinska_RHO_co2_ChemMat_2014} \cite{lozinska_RHO_co2_ChemMat_2014} have found that, once dehydrated, the monovalent forms of zeolite RHO do not change from the acentric to the centric space group when increasing the temperature. In order to get a better atomistic insight into this interesting result, we have conducted three computational experiments. The first two concern the study of the influence of water and EFCs in the structural features of zeolites \cite{B607030B,RabdelRuizSalvador20041439}. In the first study, water was initially considered as a continuum dielectric medium that screens cation-zeolite Coulombic interactions, with the aim of understanding the structural features of complex zeolites upon progressive dehydration. In the second study, water molecules were treated explicitly. Moreover, we calculated the energy barriers that the cations should surpass when traveling across the zeolite windows to go from one cationic site to another.

In order to quantify the degree of distortions we used the following parameters:
\begin{eqnarray}
\label{eq2_distortions:1}
\delta_4^t  =& \frac{1}{2}\lvert r^t_{ik}-r^t_{jl}\rvert  \\
\label{eq2_distortions:2}
\delta_6^t  =& \frac{1}{2}\lvert \max{(r^t_{il},r^t_{jm},r^t_{kn})}-\min{(r^t_{il},r^t_{jm},r^t_{kn})}\rvert \\
\label{eq2_distortions:3}
\delta_8^t  =& \frac{1}{2}\max{\left( \lvert r^t_{im}-r^t_{ko} \rvert,\lvert r^t_{jn}-r^t_{pl} \rvert\right)}
\end{eqnarray}
where $r^t_{ij}=\lVert\vec{r}_i(t)-\vec{r}_j(t)\rVert$ and $i$, $j$, $k$, $\ldots$ $o$, $p$ are oxygen atoms labeled clockwise. These parameters are defined for each window (4, 6, and 8-rings) and time $t$ (e.g. for 8-ring see Figure \ref{fig:distortion_parameter}). Thus, we calculate average values for all windows and time:
\begin{equation}
\label{parameters}
\Gamma=\langle \delta_4\rangle ,\quad\Lambda=\langle \delta_6\rangle ,\quad \Delta \equiv \langle \delta_8\rangle .
\end{equation}
Equation \ref{eq2_distortions:3} is a generalization of the degree of ellipticity of D8R, originally described as in Equation \ref{Deltadef}. This is motivated by a previous experimental work on zeolite LTA \cite{ITQ-50}, where a complex behavior for the window distortion was observed. Equation \ref{eq2_distortions:3} was defined in a previous work \cite{RHO_delta_Parise_1983}, however Equation \ref{eq2_distortions:1} and Equation \ref{eq2_distortions:2} are first introduced in the present work. A state-of-the-art algorithm was used for the automatic non-trivial identification of all window-types \cite{dominguez2014loops}, which is based on loops searching in empirical networks treating each zeolite window as a loop in a dynamic graph. In addition to the three distortion parameters ($\Gamma$, $\Lambda$, and $\Delta$), the average cell parameter allows us to monitor the geometry of the unit cell during the simulations. For computing the average values extracted from the simulations and reported below, ergodicity has been assumed, and therefore the average values are calculated as the corresponding ensemble average \cite{landau1980statistical,FrenkelSmit}. In this way, the more stable configurations have a larger stability and contribute therefore more to the computed observables.

The MC simulations have been carried out using the code RASPA \cite{RASPA_code}, while the EM and MD ones have been carried out using the GULP code \cite{GULP}. Constant pressure MD simulations with isotropic volume fluctuations and fully flexible unit cells have been used to study the evolution of the system, and to produce the correct statistical ensemble, (Nosé-Hoover thermostat with Rahman-Parrinello barostat \cite{parrinello:7182}) since it allows for phase changes in the simulation. In the MD simulations, the pressure has been set to zero. The integration time step is 0.1 fs. Each MD step consisted on a 2.5 ps equilibration simulation, followed by a 2.5 ps production run. Electrostatic interactions are calculated using the Ewald summation \cite{PP1921,DJ1983}. A cut-off radius of 12 \AAA is used for short-range interactions.

\section{Results}
In order to model a system that is comparable to those studied experimentally, we considered the inclusion of Al atoms. To do that, two crystalline configurations are set up, replacing 80 Si atoms by 80 Al atoms (per computational box, consisting of a $2\times 2\times 2$ supercell). The Al atoms are placed in such a way that first, they obey L\"{o}wenstein’s rule \cite{loewenstein}, and second, the Coulombic repulsions between Al-centered tetrahedron are minimized. The first configuration is labeled as C1, and has the Al atoms placed as far as possible in a $2\times 2\times 2$ supercell. The average distance between Al atoms is $\langle\text{Al}_i \text{Al}_j\rangle_{\text{C1}}= 14.49$, and the average distance between the closest pairs of Al atoms is $\langle\mathcal{C}(\text{Al}_i\text{Al}_j)\rangle_{\text{C1}}=5.73$~\AA. The second configuration (C2) is constructed by placing the Al atoms as far as possible in a single unit cell, which is subsequently expanded to build a $2\times 2\times 2$ supercell. The average distance between Al atoms and between closest Al pairs are $\langle\text{Al}_i \text{Al}_j\rangle_{\text{C2}}= 14.50$ \AA, and $\langle\mathcal{C}(\text{Al}_i \text{Al}_j)\rangle_{\text{C2}}=5.64$~\AA, respectively. C2 has a higher symmetry than C1, so the average distance between closest pairs of Al atoms is circa 0.1 \AA~higher. This is just a small increase, which suggests that it is not necessary to employ the larger unit cell to model the system correctly, since the small improvements that would be achieved by increasing Al-Al repulsions would be insufficient to compensate for the increase in simulation time. We will therefore use the configuration C1 for the rest of this work. The atomic coordinates of the framework atoms (Si, Al and O atoms) of both structures are provided in ESI.

\begin{figure}[htb]
	\begin{center}
	   \centering
	   \includegraphics[width=0.49\textwidth]{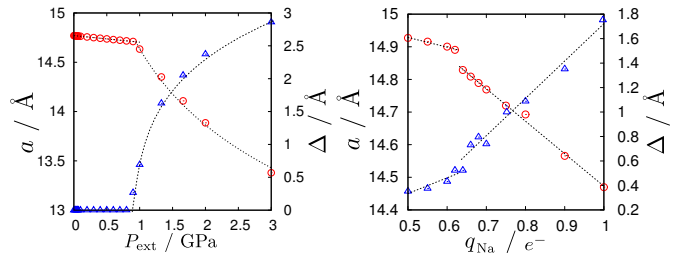}
	   \caption{
	   \label{fig:stress}
	   Left: cell parameter, $a$ ($\circ$), and 8MR window distortion $\Delta$ ({\small $\triangle$}), \vs applied external pressure. Right: cell parameter and 8MR window distortion \vs partial charge for sodium cations, $q_{Na}$. We observe a structural phase transition at 1 GPa and $q_{Na}=0.62~ e^{-}$, respectively. The errors are smaller than the size of the points. All regressions have $r^2 > 0.99$.
	   }
	\end{center}
\end{figure}

The strong attractive interaction between the extra-framework cations and the oxygen atoms of the zeolite exerts a large force on the framework that  is likely to be the cause of the phase transition from the centric \centric to the acentric \acentric structure with the concomitant reduction of the cell volume and the increase of the pore window acentricity ($\Delta$ parameter). In order to shed more light into this feature, we analyzed the behavior of the pure silica zeolite RHO in presence of an externally applied pressure. Figure \ref{fig:stress}-left shows that, indeed, framework volume decreases gradually until a step is found at 1 GPa. At this point the acentricity parameter is 0.25 \AAA, and the crystal structure adopts the acentric form. This explains the origin of the phase transition of zeolite RHO. In the remaining of the paper we will explore it to greater depths, as well as the structural features related to the cell deformation and its impact on the molecular nanovalve effects.

In a previous work it was shown that a qualitative picture of the role of the extra-framework cations in the zeolite structural deformation can be obtained by mimicking the screening effect of water by decreasing the charge of the extra-framework cations \cite{B607030B}. The curve displaying the variations of the cell parameter as a function of the charge of the cation is shown in Figure \ref{fig:stress}-right for Na-RHO. We observe again a gradual variation, followed by a step, where the acentricity parameter undergoes a large change. The symmetry also changes from the centric to the acentric space group. The decrease of the charge of the cations could be regarded as an increase in water content, and is accompanied by a reduction of cation-oxygen interactions, and consequently by larger interatomic distances \cite{B607030B}. This can cause the migration of cations, which indeed moved from Site I to Site II. 
\begin{figure}[h!t]
	   \centering
	   \includegraphics[width=0.5\textwidth]{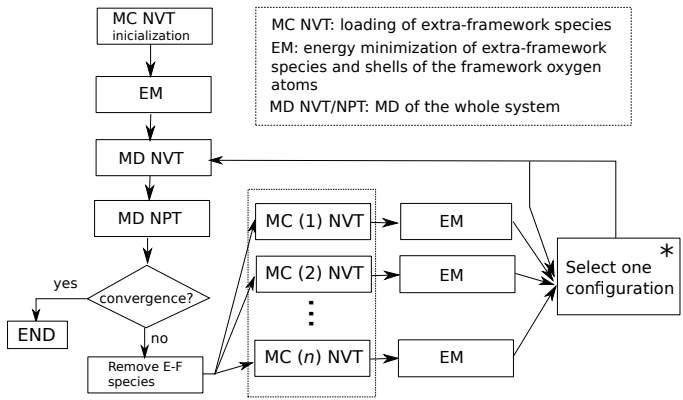}
	   \caption{\label{sch:1}EM/MC/MD simulation scheme used in this work. (*) We select the most stable configuration, according to their Boltzmann weights, among the previously accepted configuration and a new set of ten trial calculated systems.}
\end{figure}

For a realistic modeling of highly-flexible porous nanoporous solids, we have used a methodology with which to achieve a step by step approach towards the most likely structure of these complex materials. Zeolite RHO, according to experimental evidences, is among those nanoporous materials where there are many available sites for cation siting and a close coupling exists between cation siting, cell parameters and crystal symmetry. In this case, it is not possible to find an initial structural model that leads to a realistic description of the zeolite through a single-step molecular simulation method (energy minimization, Monte Carlo or Molecular Dynamics). This is a consequence of the very complex surface energy, and thus any random trial is likely to change the structure around the closest local minimum. To surmount this, we aimed to design a method that would be able to approach the structure in an adaptive way. This methodology combines MC, EM and MD in iterative cycles. A schematic view of the method is shown in Figure \ref{sch:1}. This scheme shows the main feature of the method, which is the cyclic nature of the MC/EM/MD steps. The starting point is a stable framework structure, for instance the experimental structure (either solvated or desolvated), or a hypothetical structure. Ten trial MC runs are then performed, in order to locate the extra-framework species. In the present case these extra-framework species are extra-framework cations and/or water molecules. This is followed by a lattice EM for each MC configuration, keeping the cell parameters fixed as well as the coordinates of the framework atoms. If the EM were performed allowing the cell parameters to vary, at constant pressure, the framework structure would change abruptly and it would be trapped in a local minimum with a large distortion, from which it would be very difficult to escape. At this stage the most important issue is to achieve the relaxation of the extra-framework species, to prevent unrealistic high forces at the beginning of the MD run, which would cause the collapse of the simulation.
The energies of these ten configurations along with that accepted in the previous cycle are used to select one according to Boltsmann's weight as compared with a random probability.
In this way, detailed balance is fulfilled \cite{chandler1987introduction,FrenkelSmit,landau2013course}. 

Our aim is to provide a realistic description of the material, for which thermal and entropic contributions are relevant. Since in EM calculations, these are lacking, we decided to model the overall structural changes of the solid by means of MD simulations. Then, after performing the EMs, to further ensure that the MD can be performed in a reasonable way, a short run is first accomplished at fixed cell parameters, in the NVT ensemble, followed by a long run in the NPT ensemble, with a fully flexible cell. In the later run the cell parameters vary according to the effect of the extra-framework species and external conditions like temperature and hydrostatic external pressure. Since the coupling between the framework structure and the extra-framework species is quite strong, the MD simulation relaxes the structure primarily in accordance to the location of the extra-framework species. Moreover, in this MD stage some extra-framework species can surpass the local energy barriers and move to adjacent stable cation/adsorbate sites. This movement to other cation/adsorbate sites causes in turn changes in the framework and the overall structure of the material. After each MD run, the obtained framework structure is taken for the new cycle, while the extra-framework atoms are removed and inserted back with the MC scheme step. This cyclic process is repeated iteratively, until the potential energy of the system is equilibrated.

\begin{figure}[t]
	   \centering
	   \includegraphics[width=0.48\textwidth]{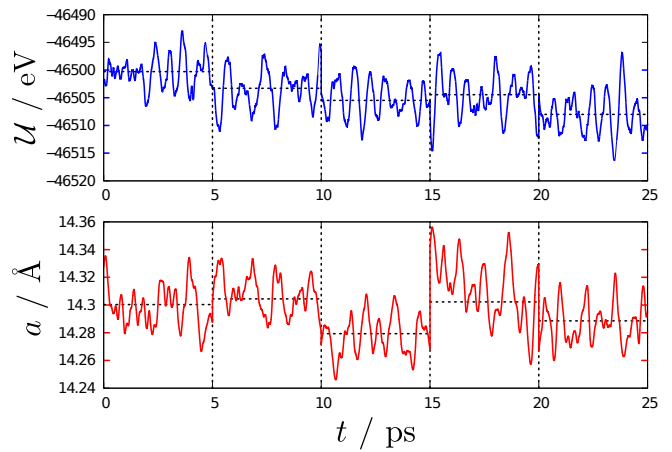}
	   \caption{\label{fig:cycles}
	   Variation of the potential energy (top) and cell size (bottom) with the number of MD simulation steps, for the first 5 MC/EM/MD cycles, for Sr-RHO. Horizontal dashed lines represent the average potential energy and cell size, respectively. Vertical dashed lines represent boundaries between cycles. The temperature is 300~K.
	   }
\end{figure}

Figure \ref{fig:cycles} shows the evolution of potential energy and cell size, as when the described method is employed to model zeolite RHO with Sr$^{2+}$ cations. For clarity reasons, only the first five MD/EM/MC cycles are shown, and the complete curve is presented in the ESI (see Figure S4). At the beginning of each MC/EM/MD-cycle there is a sudden increase in the cell parameters, resulting from the repulsion induced by the extra-framework cations located in new sites with respect to their previous positions. The response of the whole system is to slightly increase cation-zeolite distances, while at the same time the structure is relaxed. During the transit along the MC/EM/MD cycles, the internal energy, i.e. the average of total potential energy, decreases and converges.  We observe that the potential energy tends to decrease as the methodology proceeds. However the behavior of the cell size is complex, which as a consequence of the structural variation of the system trying to gain stability upon relocation of the extra-framework cations in each MC step. In fact this is the core point of our methodology.

\begin{table*}[t]
\caption{\label{cell_parameter_temp}Dependence with temperature of the average cell parameter, $a$, for silica and aluminosilicate forms of RHO-type zeolites.}
\begin{tabular*}{\textwidth}{@{\extracolsep{\fill}}cccccccc}
\\
Me--RHO$_{n_{\text{Al}}}$  &  \multicolumn{2}{c}{Na$_{80}$} &\multicolumn{2}{c}{K$_{80}$} & \multicolumn{2}{c}{Ca$_{80}$} & \ce{SiO2} \\ \hline
$T_{\text{ext.}}$ & $\langle a\rangle$ /  \AA  & \multirow{2}{*}{Exp.[Ref.]}  &$\langle a\rangle$ /  \AA  & \multirow{2}{*}{Exp.[Ref.]}  &$\langle a\rangle$ /  \AA  & \multirow{2}{*}{Exp.[Ref.]} &  $\langle a\rangle$ /  \AA \\
/ $K$ & $\pm 0.001 \AA$ & & $\pm 0.001 \AA$ & & $\pm 0.001 \AA$ & \\ \hline
100 & 14.619 &  & 14.964 & & 14.379 & & 14.757\\
200 & 14.554 &  & 14.722 & & 14.195 & & 14.635\\
\multirow{2}{*}{300} & \multirow{2}{*}{14.431} &  14.4139\text{\cite{lozinska_RHO_co2_jacs_2012}} & \multirow{2}{*}{14.602} & 14.5951\text{\cite{lozinska_RHO_co2_jacs_2012}} & 14.141 &  &\multirow{2}{*}{14.487}\\
& & 14.3771\cite{lozinska_RHO_co2_ChemMat_2014} & & 14.5959\text{\cite{lozinska_RHO_co2_ChemMat_2014}} & &  \\
350 & 14.475 & & 14.650 & & 14.004 & & 14.395\\
500 & 14.467 & & 14.613 & & collapsed & & 14.438\\
600 & 14.423 & & 14.805 & & collapsed & & \\
& \multicolumn{2}{c}{Li$_{80}$} &\multicolumn{2}{c}{Sr$_{80}$} & \multicolumn{2}{c}{Na$_{92}$} \\ \hline
100 & 14.491 & & 14.429 & & 14.340 & &\\
200 & 14.330 & & 14.388 & & 14.103 & &\\
300 & 14.208 & 14.2448\cite{lozinska_RHO_co2_jacs_2012} & 14.402 & & 14.141 & &\\
350 & 14.102 & & 14.352 &  & 14.004 & &\\
500 & collapsed & & 14.173 &  & collapsed & &\\
600 & collapsed & & 14.187 &  & collapsed & &\\
& \multicolumn{2}{c}{K$_{92}$} &\multicolumn{2}{c}{Ca$_{92}$} & \multicolumn{2}{c}{Sr$_{92}$}  \\ \hline
100 & 14.747 & & 14.340 & & 14.382 & \\
200 & 14.669 & & 14.203 & 14.489 & 14.412 &\\
300 & 14.607 & & 14.058 & 14.45\cite{RHO_divalent_cation_jacs} & 14.327 & 14.45\cite{RHO_divalent_cation_jacs} \\
\multirow{2}{*}{500} & \multirow{2}{*}{14.498} & & \multirow{2}{*}{collapsed} & \multirow{2}{*}{14.01\cite{RHO_divalent_cation_jacs}} & \multirow{2}{*}{14.259} & 14.56\text{\cite{RHO_divalent_cation_jacs}} \\
& & & & & & 14.05\text{\cite{RHO_divalent_cation_jacs}} \\
600 & 14.498 & & collapsed & 14.07\cite{RHO_divalent_cation_jacs} & 14.211 & 13.98\cite{RHO_divalent_cation_jacs} \\
\end{tabular*}
\end{table*}

We have employed the MC/EM/MD algorithm to study the dependence with temperature of the cell size, for all cations. The results obtained in our simulations, as well as the available experimental data, are reported in Table \ref{cell_parameter_temp}. We have simulated zeolites containing 80 and 92 aluminum atoms per simulation cell, with Si/Al ratios of 3.8 and 3.17, respectively, in order to compare with experiments. We observe negative thermal expansion (NTE) in all cases, but unlike what \citeauthor{RHO_relocations_cations_reisner} \cite{RHO_relocations_cations_reisner} reported, our results suggest that the negative thermal expansion is not associated to a gradual dehydration but to the intrinsic response of the dehydrated aluminosilicate to an increase of temperature. It is worth noting also that the pure silica hypothetical zeolite RHO shows NTE. In relation to this point, the collapse of the structures with high polarizing powers occurs due to the limitation of the potentials, but the main picture is well described, as the same behavior is observed experimentally but at higher temperatures. Finally, it is worth noting that the difference of the computed cell parameters with experiments is below 0.2 \%. This is the first theoretical work that achieves such a close agreement with experimental data, in a highly flexible zeolite containing Al atoms and extra-framework cations.
\begin{figure}[h]
	\begin{center}
	   \centering
	   \vspace*{0.5cm}
	   \includegraphics[width=0.45\textwidth]{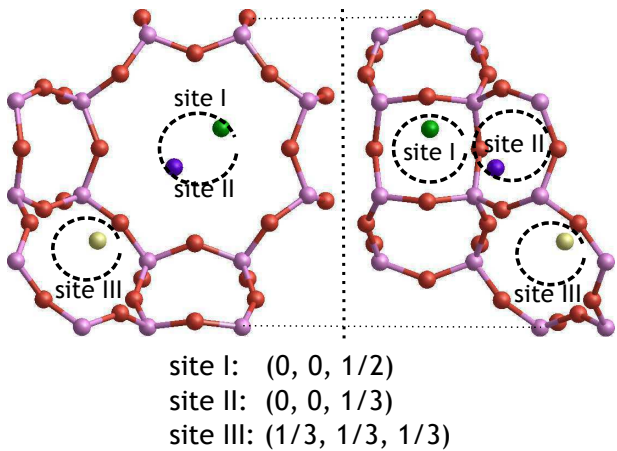}
	   \caption{\label{fig:sites}Atomistic view of the cationic sites in zeolite RHO sites I-III depicted as dashed spheres and its coordinates.        
	   The colored spheres represent cations in cationic sites.
	   }
	\end{center}
\end{figure}

In connection with the above results, here we address the EFCs location and the population of the different cationic sites, as obtained from our simulation. Experimentally three crystallographic sites are known for EFCs in zeolite RHO: (site I) inside  a double 8-ring, (site II) at the center of a single 8–ring, and (site III) in front of the 6–ring (see Figure \ref{fig:sites}). In agreement with the experimental data \cite{RHO_Co2_palomino,RHO_cation_siting,
RHO_divalent_cation_jacs,lozinska_RHO_co2_ChemMat_2014,lozinska_RHO_co2_jacs_2012}, our MC/EM/MD methodology shows that most cations are distributed among the known cationic sites, while a small fraction was assigned elsewhere. This means that the latter were located outside of a 2 \AAA radius sphere from the reference sites, which is mainly because EFCs are found moving from one stable site to another. The population of the sites is in good agreement with experimental results, when available, considering the uncertainties associated with the experimental data that can be extracted from diffraction (see Table \ref{tab:population}). Note for example that the overall cation content determined in experiments varies from 80 to 95 \% of the nominal EFCs content. The largest difference was found for Li$^+$ with 10 cations per unit cell (80 cations per simulation cell), where experiments allocated 8 of these atoms to site III, whereas we found 5.3 atoms at this site, 2.7 at site II, 1.0 at site I and 1.0 elsewhere. The sum of the populations of sites II and III matches very well the experimental value obtained for the latter site. This is located near site II and, considering the large uncertainty that can be associated with the location of Li$^+$ cations by X-Ray diffraction, we can conclude that the simulations provide a reasonable agreement with experiments. When the number of cations increases, concomitant with the increase of Al-atoms per unit cell from 10 to 11.5, we obtain that the occupation of site III increases for Na$^+$ and K$^+$, while for the case of Ca$^{2+}$ and Sr$^{2+}$ this site depopulates in favor of site I. The success in reproducing both the cell parameter behavior and the EFCs location provides confidence in the ability of our calculations to correctly describe the complex behavior of zeolite RHO and analyze in detail the dynamical features of the pore window, and its connection to nanovalve effects.

\begin{table}[h]
\small
\caption{Cation site populations in atoms / unit cell within a 2 \AA~radius from the cation sites at temperatures 300~K.\textsuperscript{\emph{a}}}
\label{tab:population}
\begin{tabular*}{0.5\textwidth}{@{\extracolsep{\fill}}lllll}
\hline
cation & site I & site II & site III & elsewhere
\\    
Na$_{80}$  & 1.64 & 6.55 & 1.61 & 0.08 \\
Exp. \cite{lozinska_RHO_co2_ChemMat_2014} & & 6.48 & 3.00 & \\ \hline
Na$_{92}$ & 1.36  & 7.09 & 2.94  & 0.07 \\ \hline
K$_{80}$ & 2.06 & 5.54 & 2.36 & 0.09 \\
Exp. \cite{lozinska_RHO_co2_ChemMat_2014}  & 2.06 & 6.85 & & \\ \hline
K$_{92}$ & 2.08 & 5.91 & 3.39 & 0.10 \\ \hline
Li$_{80}$ & 1.04 & 2.68 & 5.28 & 1.03 \\
Exp. \cite{lozinska_RHO_co2_ChemMat_2014}  & & & 8.00 & \\ \hline
Li$_{92}$ & 1.30 & 3.49 & 5.38 & 1.37 \\
Ca$_{80}$ & 2.00 & 1.99 & 0.75 & 0.26 \\
Ca$_{92}$ & 3.36 & 2.06 & 0.21 & 0.13 \\
Sr$_{80}$ & 1.91 & 2.24 & 0.74 & 0.11 \\
Sr$_{92}$ & 3.01 & 2.45 & 0.27 & 0.02 \\
\hline
\end{tabular*}
\begin{flushleft}
\textsuperscript{\emph{a}} Cation sites from: \cite{RHO_Co2_palomino,RHO_cation_siting,RHO_dehydration_jpc}
\end{flushleft}
\end{table}

Since the methodology does not impose any symmetry constraints other than the supercell, it is able to provide structural and dynamical data that are close to the behavior of the materials. This can be exploited for increasing our understanding of this complex behavior, in particular in materials with a large degree of flexibility, such as zeolite RHO. Previous simulation work \cite{awaiti2013,Sastre201425} has shown that, even for rather rigid zeolites, the size of the windows that delimit the pore opening has large deviations (circa 0.4 \AA), due to the thermal motion. Diffraction studies cannot capture this feature accurately due to the symmetry-averaged picture it provides, even by analyzing the information extracted from the Debye-Waller factors. In the present context, it is useful to note that despite the role that the flexibility of the windows plays in the transport properties of zeolites without EFCs \cite{awaiti2013,Sastre201425}, how the windows flexibility is modified in the presence of EFCs remains unexplored. As we have shown above, we have confidence in the accuracy of the developed computational scheme to provide accurate structural data and therefore we can enter into the detailed analysis of the window deformations for the different metal-forms of zeolite RHO.
\begin{figure}[th]
	\begin{center}
	   \centering
	   \vspace*{0.5cm}
	   \includegraphics[width=0.42\textwidth]{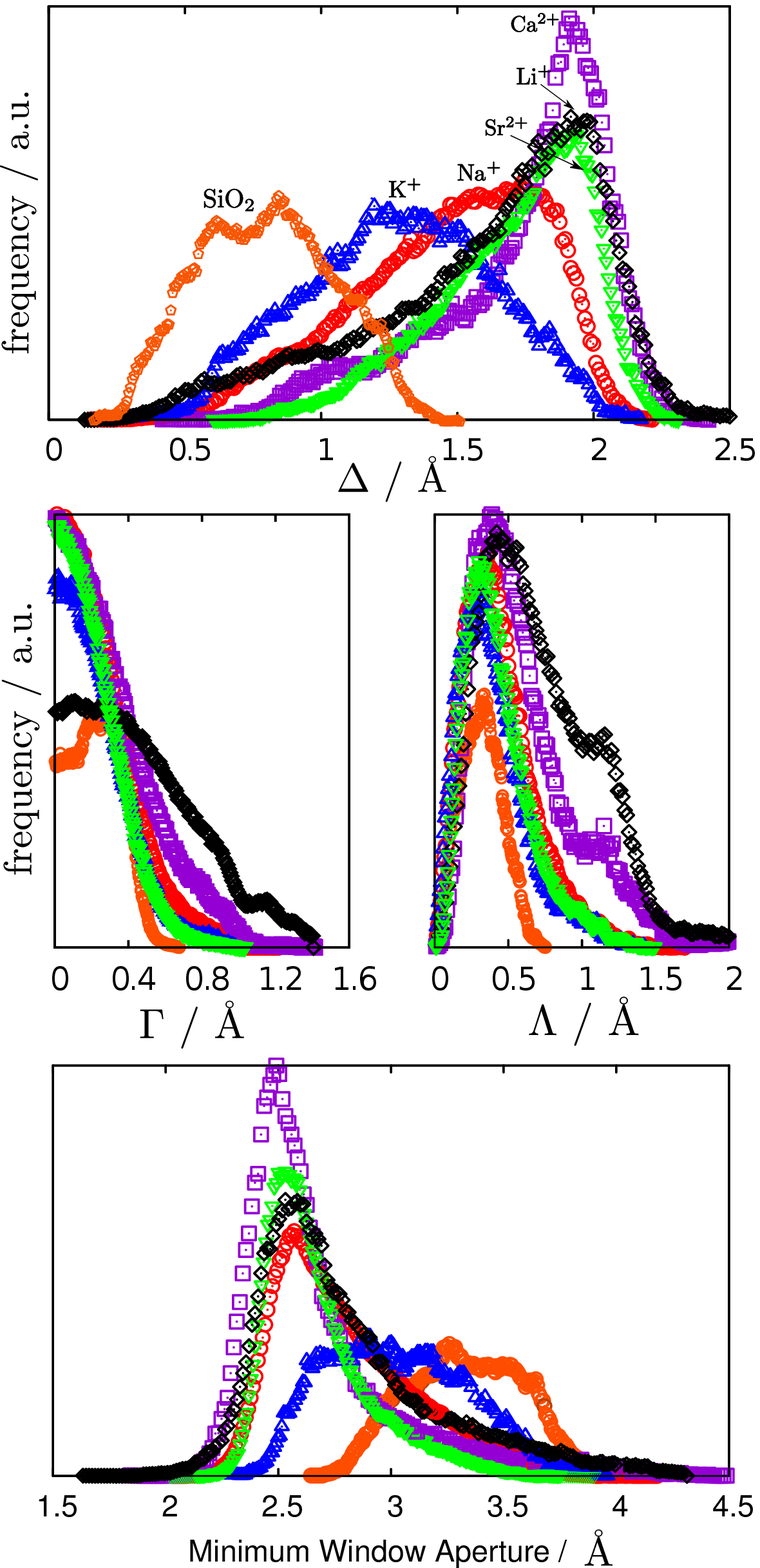}
	   \caption{\label{fig:distortions}
	   (Top and middle) Probability density of degree of distortion of the 4,6,8-rings ($\Gamma$, $\Lambda$, and $\Delta$) for pure silica and aluminosilicate forms (Li$^{+}$, Na$^+$, K$^+$, Ca$^{2+}$, and Sr$^{2+}$). (Bottom) Probability density of minimum window aperture of the 8-rings aluminosilicate forms (Li$^{+}$, Na$^+$, K$^+$, Ca$^{2+}$, and Sr$^{2+}$). The temperature is 300~K in both figures.
	   }
	\end{center}
\end{figure}

Figure \ref{fig:distortions} shows the window distortion profiles for pure silica and aluminosilicate forms (Li$^+$, Na$^+$, K$^+$, Ca$^{2+}$ and Sr$^{2+}$), by displaying the deformation parameters defined for the 4,6,8-rings ($\Gamma$, $\Lambda$, and $\Delta$). It is worth noting that the pure silica structure shows non-zero window distortion, which is due to the thermal motion and is a fact that cannot be directly inferred by diffraction methods. We have performed \textit{Ab initio} Molecular Dynamic simulations that support these findings; details of which are provided in ESI. The distortions observed for this case can be regarded as those intrinsic to the topology and thus the deviations from them observed for the metal forms can be interpreted as being induced by the cations. 4-membered rings (4MRs) are kept mainly undistorted ($\Gamma$ maximum population corresponds to $\Gamma =0$ \AA, except for pure silica structures), due to their small size, although they exhibit relatively long tails, particularly Li$^+$ and Ca$^{2+}$. Obviously, 4MRs do not directly contribute to molecular transport, but their deformation can couple to the deformation of larger windows, and could enhance the pore opening of these ones.
The distortions of the 6MRs are not centered at zero, but circa at 0.5 \AAA for all forms, and they show a small dependence on the nature of the cations. This is also due to the small size of this type of window, although they have noticeably larger tails as well as a shoulder at around 1 \AAA for Li$^+$ and Ca$^{2+}$. In the case of 8MRs a clear dependence of the $\Delta$ parameter with the metal polarizing power is observed. Note that for this distortion there is a much larger departure from the behavior of the pure silica zeolite. Going back to the smaller rings, \textit{i.e.} 4 and 6MRs, it can be observed that indeed the effect of the polarizing power
in these cases.
These findings are consistent with the idea that distortions are coupled and cations produce distortions in all rings.
On this basis it is possible to understand one interesting result of \citeauthor{lozinska_RHO_co2_jacs_2012} \cite{lozinska_RHO_co2_jacs_2012} who found, in their study of monovalent cation exchanged zeolite RHO, that despite not being located next to 8MR, the Li$^+$ cation is the one that induces the largest values of the mean $\Delta$ parameter. On the other hand, a direct relation between cation site occupancy and framework distortion has been found in divalent cation-exchanged zeolite RHO by \citeauthor{RHO_divalent_cation_jacs} \cite{RHO_divalent_cation_jacs}.

\begin{figure}
	\begin{center}
	   \centering
	   \vspace*{0.5cm}
	   \includegraphics[width=0.42\textwidth]{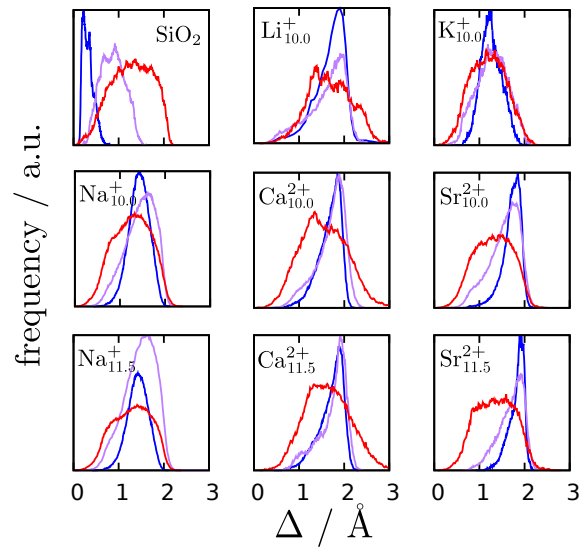}
	   \caption{\label{fig:distortions-temperature}
	   Probability density of degree of distortion of 8-rings for pure silica and aluminosilicate forms (Li$^{+}$, Na$^+$, K$^+$, Ca$^{2+}$, and Sr$^{2+}$) at different temperatures 100, 300 and 500~K (blue, purple and red, respectively) and Si/Al ratios (3.8 and 3.17).	   }
	\end{center}
\end{figure}
\begin{figure*}[thb]
	\begin{center}
	   \centering
	   \includegraphics[width=0.65\textwidth]{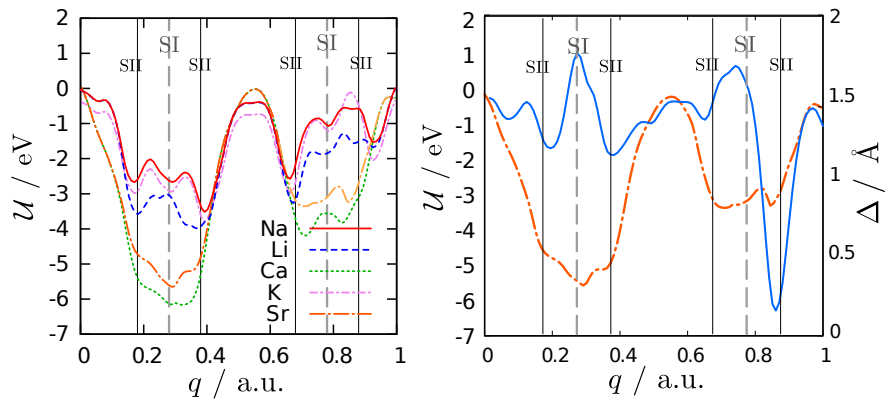}
	   \caption{\label{fig:free-energy}
	   (Left) Potential energy landscape $\mathcal{U}$ \vs reaction coordinate (in the $a$-direction through the double 8--rings), $q$, for one or two aluminum atoms per unit cell. The position of the first aluminum atom is $q = 0.37$ and the second (for the divalent cases) is $q=0.17$. (Right) Potential energy landscape $\mathcal{U}$ and distortion parameter $\Delta$ \vs $q$ reaction coordinate for Sr-RHO form.
	   	   }
	\end{center}
\end{figure*}

The behavior of the distortions in 8MRs with temperature is dependent on the nature of the metal cation present in the zeolite, in a very distinct manner. For each structure with divalent cations the distortions are about the same at 100~K and 300~K, while at 500~K the position of the maximum of the curves are reduced circa 0.5 \AAA and the peak widths are doubled (Figure \ref{fig:distortions-temperature}). The increase of the number of cations for these two metals has almost a negligible effect on the observed behavior, which suggests that even at the lower number of cations, their interactions with the zeolite oxygen atoms are very strong. For the monovalent cations the influence of both the polarizing power and the number of cations is clear. Li$^+$ has the smallest cationic radius and thus the largest polarizing power, and shows only a minor shift of the position of the maximum of the degree of distortion. The behavior of the Na$^+$ form shows variations with the number of cations and also with temperature. From 100~K to 300 K the distortions are larger, while they decrease at 500~K. In the case of K$^+$, which has the largest radius and lowest polarizing power, very little variation of the maximum of the degree of distortion are observed, even at 500~K. As was more clearly shown in Figure \ref{fig:distortions}-top, the positions of the peaks of the distortion distributions of the three metals with largest polarizing power (Ca$^{2+}$, Li$^+$ and Sr$^{2+}$) do not differ in a noticeable fashion, which together with the information gained from the analysis of the temperature dependence suggests that the largest window distortion that can occur in zeolite RHO under experimental conditions is about 2 \AA. The windows of pure silica zeolites exhibit increasing distortions with an increase in temperature. The overall analysis of Figure \ref{fig:distortions-temperature} reveals that at 500~K the thermal effect makes the behavior of the window distortion independent of the nature of the cation and more dictated by the framework itself.

In agreement with \citeauthor{lozinska_RHO_co2_jacs_2012} \cite{lozinska_RHO_co2_jacs_2012}, for the dehydrated metal containing zeolites we do not find a phase transition from the acentric \acentric to the centric \centric form when increasing temperature. However, one would expect that such a phase transition might occur. In order to shed more light on the role of EFCs affecting the structural behavior of RHO zeolites, we analyzed the energy barrier for crossing an 8MR, by mapping the energy profile of the cations moving along a path traversing the zeolite following a line that crosses the 8MR (see Figure \ref{fig:free-energy}) and allowing the structure to relax \textit{via} EM. The data in the middle of the zeolite has no physical meaning, as cations will not get into  these sites in absence of adsorbed molecules, they will rather move along the pore surface. The attention should be focused on the regions close to the zeolite windows, identified in the figure by vertical continuous lines. The centers of the D8R are represented by vertical dashed lines. We observe that the energy barriers are asymmetric, \textit{i.e.} they are different for forward jumps and backward jumps, respect for D8R.  Moreover, the relative well depth depends on the location of the Al in each case. The position of the first cation, for monovalent forms, is $q=0.37$. For divalent forms a second cation is added at $q=0.17$.

An important conclusion we can draw from the analysis of the values of the energy barriers is that, in the dehydrated state, the cations need to overcome very high energy barriers in order to jump from one site to another, suggesting that phase transitions induced by cation jumps will take very long times.
This is a plausible explanation of the experimental observations of 
\citeauthor{RHO_simulation_temperature_Sr} \cite{RHO_simulation_temperature_Sr} who observed that the high temperature structure Sr-RHO quenched to 90~K, and left in vacuum at room temperature, takes a week to recover the usual room temperature structure. It is interesting to note that the overall structure responds to the cation reaction coordinate, \textit{i.e.} when the cation is in site I of the 8-ring closest to the cation ($q=0.27$) the degree of distortion is $\delta_8=1.75$ (Equation \ref{eq2_distortions:3}). However, when the cation is in site II $\delta_8=1.3$. We have not measured the degree of ellipticity induced by cations in site III. For the second D8R,  $\delta_8$ is noticeably low, at less than 0.5 for a reaction coordinate of $q=0.85$. This indicates that the local geometry is not so much affected by the presence of the cation as it is by the simultaneous presence of the extra-framework cation and the aluminum atom of the framework. Our result could also provide a rationalization for the experimental finding that, upon \ce{CO2} adsorption in Cs-RHO, NaCs-RHO and K-RHO, two crystal phases appear over a considerable range of gas pressures, which is likely to be a consequence of the trapping of cations in long-lived sites \cite{lozinska_RHO_co2_ChemMat_2014}.

\begin{figure}[!t]
	\begin{center}
	   \centering
	   \includegraphics[width=0.43\textwidth]{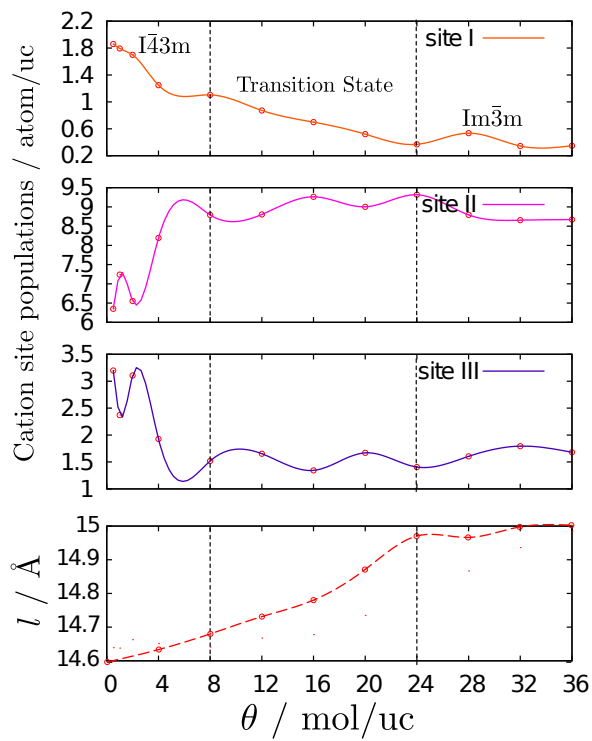}
	   \caption{\label{fig:na-populations-water}
	   Top three panels: sodium site populations (at 300~K) \vs loading of water, $\theta$, for Na-RHO with 92 Na cations. Sodium site populations are measured in atoms per unit cell, within a 2 \AAA radius from the cation sites. Bottom panel: cell size \vs water loading of Na-RHO with 92 Na cations, at 300~K.
	   }
	\end{center}
\end{figure}
\begin{figure}
	\begin{center}
	   \centering
	   \includegraphics[width=0.42\textwidth]{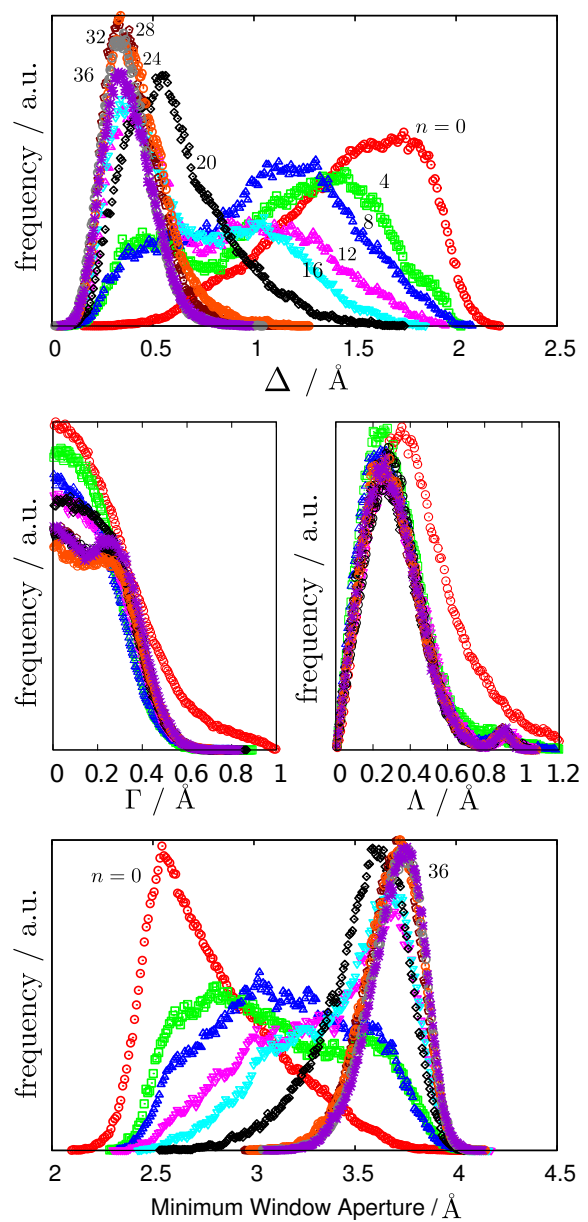}
	   \caption{\label{fig:distortions-water}
	   (Top and central) Probability density of the degree of distortion of the 4,6,8-rings ($\Gamma$, $\Lambda $, and $\Delta$) in presence of various amounts of water molecules and 92 Na cations, at 300~K. (Bottom) Probability density of minimum aperture of the 8-rings in the Na-RHO form in the same conditions.}
	\end{center}
\end{figure}
In presence of molecules that have electrostatic interactions with the zeolite, like \ce{H2O} and \ce{CO2}, it has been experimentally found that zeolite RHO can undergo reversible phase transitions \cite{RHO_Co2_palomino,RHO_dehydration_jpc}. In order to analyze how this process occurs, we have studied the behavior of Na-RHO for 8 different water contents, from fully hydrated (36 molecules per unit cell) to completely dehydrated. We started from the fully hydrated structure and the appropriate number of water molecules was eliminated for each case. Then, the systems were equilibrated by MD simulations. The lattice parameter of the initially completely hydrated material remains almost unchanged by removing up to one quarter of the water molecules, reaching 24 molecules per unit cell (Figure \ref{fig:na-populations-water}-bottom). Further gradual dehydration is accompanied by a concomitant decrease of the cell parameter, until all water molecules are removed. Our simulations show that the variations of the cell parameters are related to variations of the population of the cationic sites (Figure \ref{fig:na-populations-water}). A remarkable result is that, at any water loading, the larger amount of cations are found in site II, whose occupation is at least two times larger than those of the other two sites. Three regions are clearly identified according to the behavior of cation site population. One corresponds to high water loading, above 24 molecules per unit cell. Another region appears at low water loadings, between zero and 8 molecules per unit cell. Finally, an intermediate region is identified in between the other two. Sites I and III reach minimum occupations in the high water content region. In this case, site II is largely populated. In the intermediate region sites II and III largely maintain their occupancies, which are similar to those in the high water content region. In contrast, site I shows a gradual decrease in the intermediate region. In the absence of water, sites I and III have the largest occupation of Na cations, while the lowest occupation is observed for site II. Upon water adsorption, after the incorporation of 4 molecules per unit cell, the cationic site occupancies change rapidly and at 8 molecules, sites II and III already reach the levels that are kept about constant up to the full water loading. The observed changes of cell parameters and cationic site populations modify the symmetry of the zeolite, which has the centric \centric space group for 24 water molecules per cell up to full hydration, while adopting the acentric \acentric space group for 8 water molecules and below.  

\begin{figure}
	\begin{center}
	   \centering
	   \includegraphics[width=0.46\textwidth]{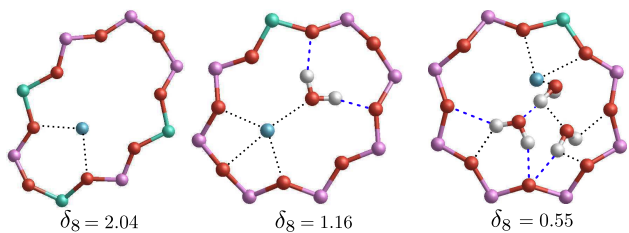}
	   \caption{\label{fig:d8r-water}
	   Snapshots of several 8MRs during simulations, when occupied by one Na$^+$ cation (left), one Na$^+$ cation and one  water molecule (middle) and one Na$^+$ cation and three water molecules (right). Note the gradual decrease of the instantaneous distortion parameter upon water presence. Blue dashed lines represent hydrogen bonds and black-dotted lines represent other strong interactions.
	   }
	\end{center}
\end{figure}

The analysis of Figures \ref{fig:na-populations-water} and \ref{fig:distortions-water} provides a more accurate view of the connection between changes in cell parameters and window distortions than that obtained earlier in this work in which water was modeled as a continuous dielectric (Figure \ref{fig:stress}-right). Figure \ref{fig:distortions-water} shows that the distortions of the smaller 4 and 6MRs are almost insensitive to the amount of water present, and are slightly lower than those of the fully dehydrated structure. In contrast, a very large dependence of the 8MR distortion parameter, $\Delta$, with the amount of water is observed for the interval delimited by 4 and 48 water molecules per unit cell (Figure \ref{fig:distortions-water}-top). From 24 water molecules per unit cell to full hydration (36 water molecules) the distortion of 8MR does not show significant variations. This is in line with the behavior of the cell parameter and the population of the cationic sites (Figure \ref{fig:na-populations-water}). Between 4 and 16 water molecules per unit cell the distribution of the $\Delta$ distortion parameter is bimodal, suggesting that some pores are closed whereas others are open. The appearance of the double peak is due to the nucleation of  water in specific sites, which is motivated by strong water-water interactions. This behavior could not be predicted by the dielectric continuum model of water in Figure \ref{fig:stress}-right.

The structural changes occurring in an 8MR when a Na$^+$ cation and varying amounts of water are presented is shown in Figure \ref{fig:d8r-water}. We observe the large distortion of the 8MR when just one cation is present, which favors the stability of the acentric structure. The addition of water molecules induces a gradual opening of the window, leading eventually to the experimentally observed transitions to the centric form.

\section{Conclusions}
A detailed investigation has been conducted in order to reach a better understanding of the distortion mechanisms that take place in molecular valve zeolite RHO. To the best of our knowledge, this is the first study that has addressed the behavior of zeolite molecular valves from a theoretical point of view. For this purpose, we have developed a new methodology based on cycles of Monte Carlo calculations, Energy Minimization and Molecular Dynamics simulations to study both monovalent and divalent cation-containing zeolite RHO, and its evolution upon changes on the nature of extra-framework cations, temperature and adsorbed molecules. The explicit consideration of the polarizability of the oxygen atoms has been found to be necessary, and it has been taken into account by using the shell model. 

The analysis of the ring distortions shows that there is a close relationship between the flexibility of the zeolite framework and the location of the extra-framework cations, as well as the water molecules, thus providing an atomistic insight that goes beyond the experimental information obtained by diffraction techniques, as no symmetry restrictions are considered. This finding is likely to have an influence on the further understanding of diffusion and separation processes, which are controlled by the molecular valve effect arising from the windows distortions. Previous theoretical works have stressed the relation between the effective diameter of the windows and the flexibility of the framework, in pure silica zeolites, but how the framework is affected by the presence of extra-framework cations had remained unexplored so far, being thus this work also pioneer in this respect. The analysis of the results obtained in our study enables us to draw the following conclusions concerning the structural features: 
\emph{a}) the phase transition from the centric to the acentric form of zeolite RHO is due to the force exerted by the extra-framework cations, which  has a similar effect to applying an external pressure,
\emph{b}) the newly developed method is accurate enough to provide the cell parameters within 0.07-0.2 \% with respect to experimental values, 
\emph{c}) the cation sites are not only well located by our computational approach but their populations are also in agreement with experiments,
\emph{d}) a clear dependence on the polarizing power of the extra-framework cation has been found for the distortions of the 8MRs,
as well as the minimun aperture of these rings that control the nanovalve effect,
\emph{e}) the distortions of 6MR and 8MR are connected, which explains the experimental observation that the largest distortion occurs for Li$^+$,
\emph{f}) the 8MR distortions follow a different behavior with temperature, depending on the nature of the extra-framework cation,
\emph{g}) cations face large energy barriers in order to pass from one site to another when the zeolite is dehydrated, being particularly high in the case of divalent cations,
\emph{h}) the amount of water present in the zeolite controls the population of cationic sites, the size of the cell parameters and the symmetry of the zeolite, and
\emph{i}) the 8MR windows distortion and pore aperture can be systematically controlled by the loading of water, allowing a fine control of the nanovalve effect.

The calculated high energy barriers for cation hopping provide a rationalization of the experimental finding that, upon \ce{CO2} adsorption in Cs-, NaCs- and K-RHO, two crystal phases appear over a considerable range of gas pressures, which according to our results is a consequence of the trapping of sets of cations in a range of configurations that lead to long-lived metastable structures. We have shown that polar molecules, in this case water, screen the large electrostatic interaction providing low energy paths for the structural change.

\section{Acknowledgments}
This work was supported by the European Research Council through an ERC Starting Grant (Sofía Calero, ERC2011-StG-279520-RASPA), the Spanish Ministerio de Economía y Competitividad (Sofía Calero, CTQ2010-16077 and CTQ2013-48396-P), and the Netherlands Council for Chemical Sciences (NWO/CW) through a VIDI grant (David Dubbeldam).

\appendix
\section{TOC entry}
\begin{figure}[h!]
	\begin{center}
	   \centering
	   \includegraphics[width=0.46\textwidth]{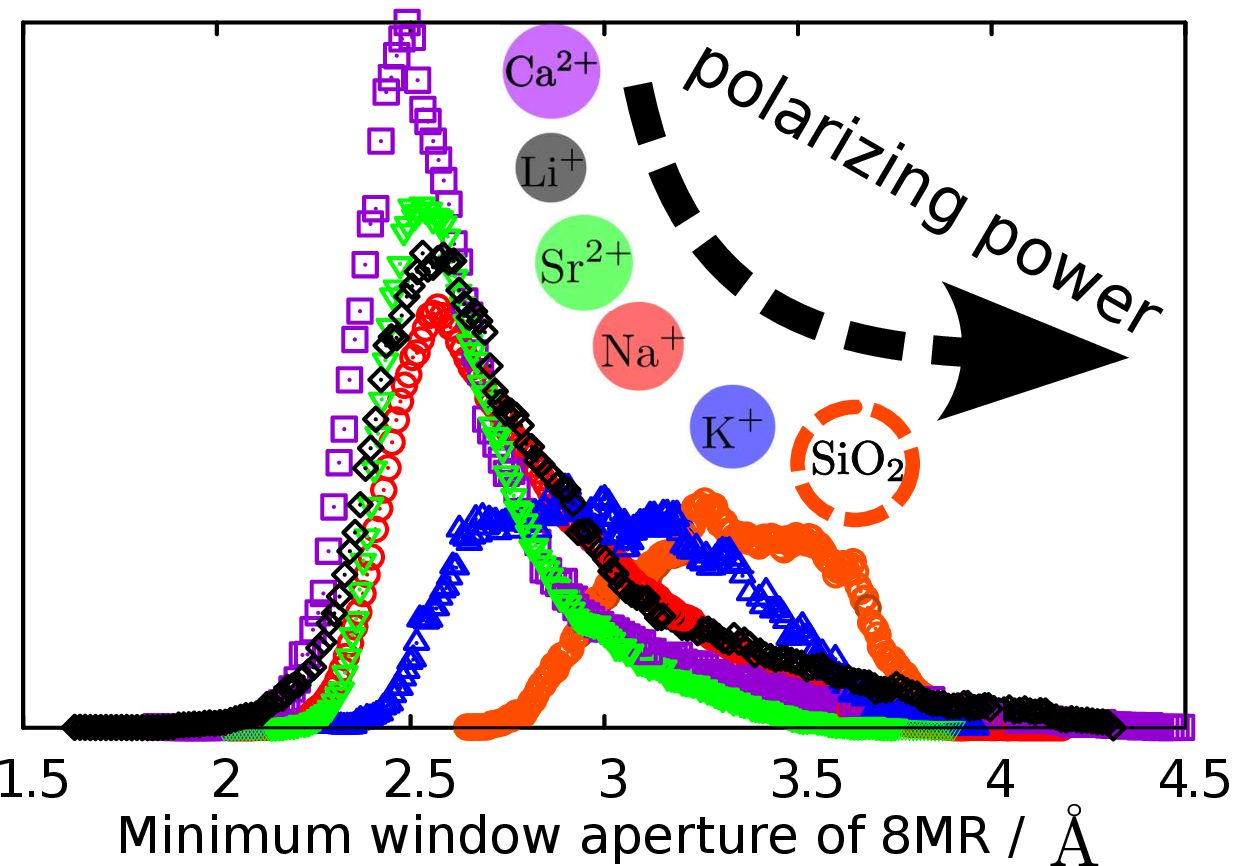}
	\end{center}
\end{figure}
\bibliography{biblio}
\end{document}